\documentclass[
 reprint, 
amsmath,
amssymb,
aps, 
pre,
]{revtex4-2}

\usepackage{graphicx}
\usepackage{dcolumn}
\usepackage{bm}
\usepackage{subcaption}
\usepackage[format=plain,justification=centering]{caption}
\usepackage{hyperref}
\usepackage{amsmath}

\begin{document}

\preprint{PREPRINT GOES HERE}

\title{\textbf{Molecular Dynamics Simulations of Bubble Nucleation in a Liquid-Noble Scintillator} 
}%

\author{Jack Walker}
 \email{Contact author: 20jdww@queensu.ca}
\author{Emma Wallace}
  \email{Current affiliation: Department of Physics and Astronomy, University of Hawai'i at Manoa, Honolulu, Hawaii, USA}
\author{Ken Clark}%
\author{Greg van Anders}
\affiliation{Department of Physics, Engineering Physics, and Astronomy, Queen's University, Kingston, Ontario, Canada.}

\author{Alex Wright}
\affiliation{Institute of Particle Physics and Department of Physics, Engineering Physics, and Astronomy, Queen's University, Kingston, Ontario, Canada.}

\date{\today}

\begin{abstract}

The Scintillating Bubble Chamber collaboration is searching for Weakly Interacting Massive Particles using a novel bubble chamber with intended thresholds as low as 100~eV. Existing molecular dynamics simulations of bubble formation in bubble chambers were conducted with non-scintillating target materials and therefore do not account for the energy transfer to photons or time-delayed releases that occur in atomic de-excitation. In this study, we use the HOOMD-blue molecular dynamics framework to simulate bubble formation in liquid argon, including photon creation, ionization, and direct nuclear recoils. A multi-stage bubble growth process similar to that reported in the literature was observed. When comparing simulated thresholds with and without scintillation effects, we found that scintillation raises the average energy required to form a bubble by a factor of 2.16. This is larger than the fraction of energy lost to photon creation, and demonstrates that energy stored in excited molecular states with lifetimes longer than the rapid growth phase of nucleation ($\sim$250 ps) does not contribute significantly to bubble formation. This conclusion was further supported by simulations showing increased bubble nucleation thresholds when the excited molecular state lifetimes were increased, even under identical thermodynamic conditions.

\end{abstract}

\maketitle


\section{\label{sec:Intro}Introduction}

Bubble chambers have been used in the search for Weakly Interacting Massive Particles (WIMPs), one of the favored candidates for dark matter \cite{Clark:2024nbx}. They consist of an active volume of superheated fluid in which  energetic ionizing particles can boil localized regions of the fluid. These bubbles can be detected optically using cameras and acoustically by piezoelectric sensors mounted around the outside of the volume capturing the pressure wave from bubble formation \cite{Snowmass_2021}. The Scintillating Bubble Chamber (SBC) collaboration is developing detectors using a volume of liquid argon (LAr) doped with xenon as the active volume. LAr is an efficient scintillator, and detection of the scintillation light can be used  to identify background events \cite{Giampa_SBC_2021, Alfonso-Pita_SBC_2023}. Importantly, scintillation also suppresses bubble formation by electron recoils relative to nuclear recoils (the expected signal from a WIMP interaction), as a larger fraction of energy is emitted as scintillation light in electron recoils than nuclear recoils. This reduces the otherwise limiting background from beta decays and gamma-ray interactions and is expected to allow SBC to be sensitive to lower energy nuclear recoil events, and hence lower mass WIMPs, than non-scintillating bubble chambers. 

Molecular dynamics simulations can be used to simulate phase transitions, including bubble nucleation, on an atom-by-atom basis using scalable cluster computing resources \cite{Holyst2010, Diemand_homogNuc_2014, Denzel_MD_2018}. Nucleation thresholds in bubble detectors are often estimated using the Seitz ``heat spike" model \cite{Seitz_heatspike_1958, PICO2}. However, this model does not account for the more complex mechanisms of energy deposition present in LAr \cite{Grosjean_chapter_1993, Grosjean_solidAr_1997, Segreto_liquidscintillation_2020}. A strong understanding of how scintillation impacts nucleation thresholds is desirable for the SBC experiment given its low intended thresholds \cite{Li_Piro_2024}. This work aims to develop a molecular dynamics simulation of bubble nucleation in a superheated fluid including the effects of scintillation, in order to assess the impact of scintillation on nucleation thresholds in an LAr bubble chamber \cite{Denzel_MD_2018}.

\section{\label{sec:Scintillation in LAr}Scintillation in Liquid Argon}
\begin{figure}[h]
    \centering
    \includegraphics[width=0.9\linewidth]{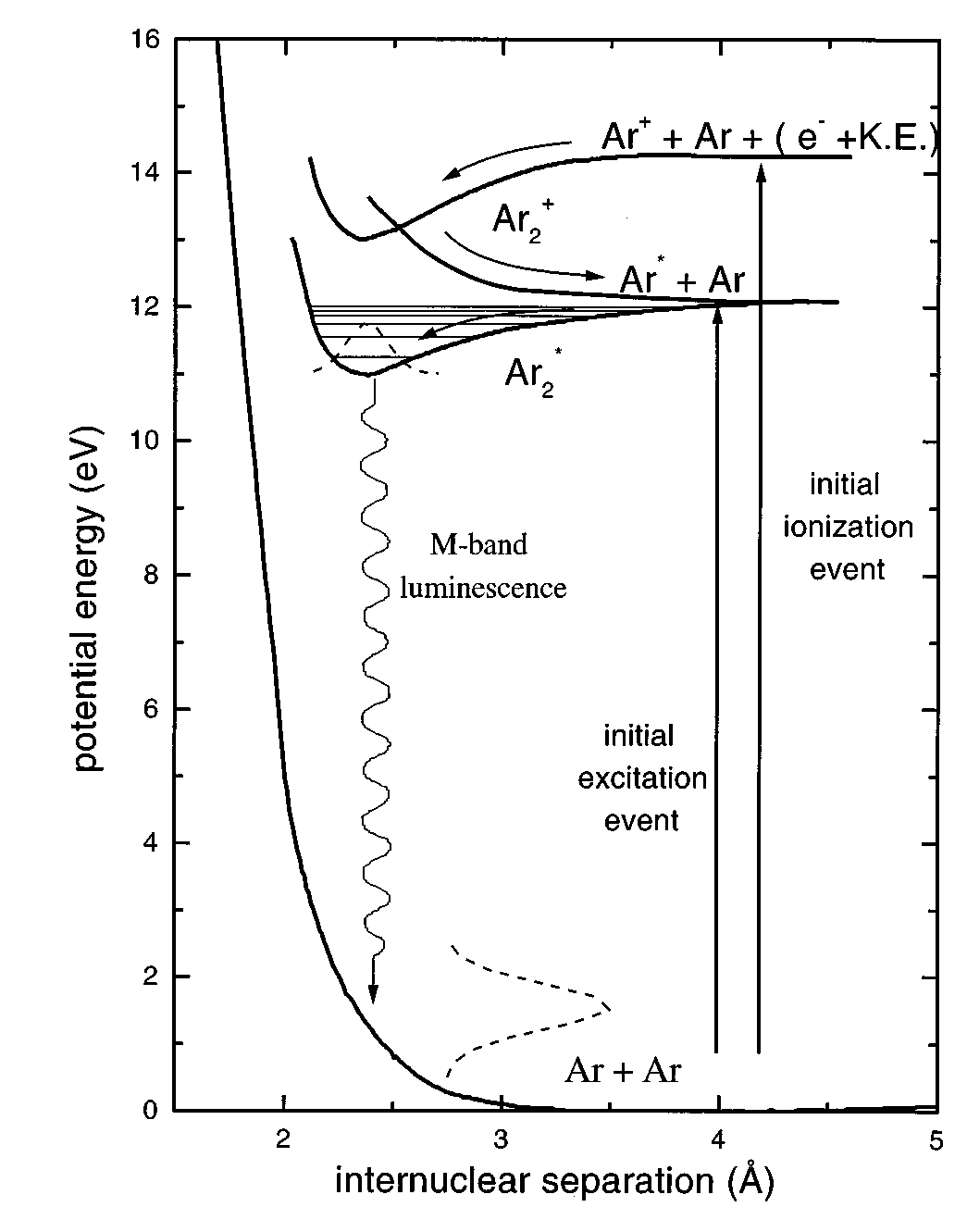}
    \captionsetup{justification=raggedright, singlelinecheck=false}
    \caption{Ar-Ar de-excitation pathways of ionized and excited atoms in solid argon. Figure from \cite{Grosjean_solidAr_1997}.}
    \label{fig:SAr_potentialdiagram}
\end{figure}

Scintillation in LAr occurs when an incident particle excites or ionizes a ground-state argon atom. When this occurs, there are two primary pathways for de-excitation, both culminating in the emission of a 9.8-eV-photon \cite{Grosjean_solidAr_1997}. These pathways are well-described in \cite{Grosjean_solidAr_1997} with respect to solid argon, and the underlying mechanisms are the same in the liquid phase; these pathways are illustrated in Figure \ref{fig:SAr_potentialdiagram}.

In cases of ionization, the incident particle frees a bound electron, creating an Ar$^+$ ion. This Ar$^+$ will combine with a ground-state argon atom to form an Ar$_2^+$ dimer, taking on the order of 1--10ps to do so \cite{Grosjean_chapter_1993}. Once the free electrons have slowed down (in approximately 100 ps), they recombine with the Ar$_2^+$ dimer to form an electronically excited argon atom, Ar$^*$, and a ground state argon atom. These argon atoms then recombine (again in 1--10 ps) to form an excited neutral argon dimer Ar$^*_2$ \cite{Grosjean_chapter_1993}. Finally, this excited dimer will decay via the emission of a 9.8 eV photon to the ground state in a time which is dependent on the electronic spin configuration; fast singlet ($\Sigma^1$) states will decay in time $\tau_f$, which is on the order of nanoseconds, while the slower triplet ($\Sigma^3$) state will decay in time $\tau_s$ on the order of microseconds.

In direct excitation, the incident particle does not deposit enough energy to free an electron, and instead produces a single Ar$^*$, skipping the first two stages in the relaxation process. The relative frequency of ionization and direct excitation, $N_{ex}/N_i = 0.21$ for energy deposition by an electron, is generally taken to be independent of energy of the incident particle \cite{Segreto_liquidscintillation_2020, DOKE_fanofactors_1976}. This ratio indicates that in a scintillation event, on average 17\% of photons result from direct excitations and 83\% from ion-hole pairs. These two processes can be summarized as follows \cite{Doke_absoluteyields_2002}:\\

Ionization (Ar$^+$):
\begin{eqnarray}
    \mathrm{Ar}^++\mathrm{Ar}\rightarrow \mathrm{Ar}^+_2 +Q_\mathrm{ion}\\
    \mathrm{Ar}_2^++\mathrm{e}^-\rightarrow \mathrm{Ar}^{*}+\mathrm{Ar}+Q_\mathrm{recomb}\\
    \mathrm{Ar}^*+\mathrm{Ar}\rightarrow \mathrm{Ar}^*_2+Q_\mathrm{excite}\\
    \mathrm{Ar}^*_2\rightarrow \mathrm{Ar} +\mathrm{Ar}+\gamma
\end{eqnarray}

Direct Excitation ($Ar^*$):
\begin{eqnarray}
    Ar^*+Ar\rightarrow Ar^*_2+Q_{excite}\\
    Ar^*_2\rightarrow Ar +Ar+\gamma
\end{eqnarray}
Here the $Q$ values represent small amounts of heat deposited in the fluid by the changes in Ar--Ar internuclear separation. This heat is generally negligible when discussing detectors with thresholds in the keV range, but with SBC expecting to operate at values as low as 100 eV, it may be relevant.

\begin{figure}[h]
    \centering
    \includegraphics[width=1\linewidth]{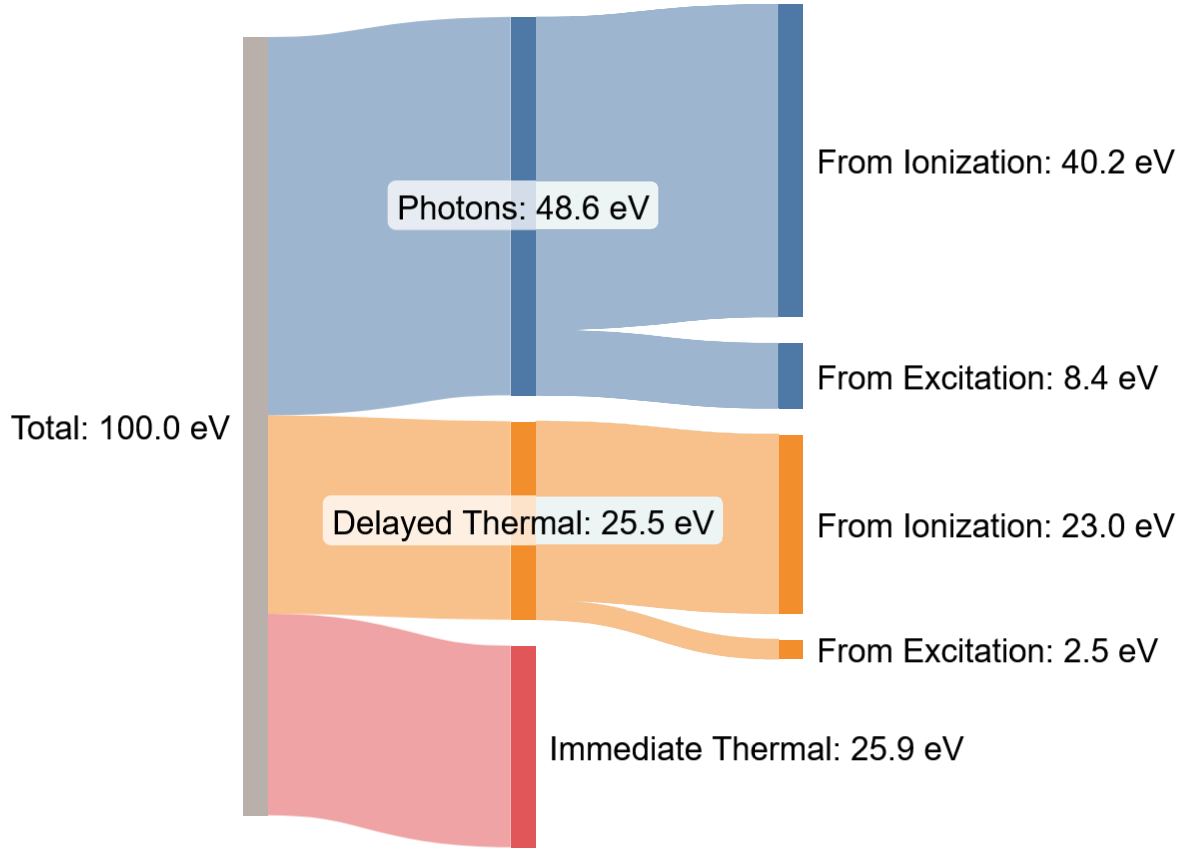}
    \captionsetup{justification=raggedright, singlelinecheck=false}
    \caption{Proportional breakdown of energy deposition mechanisms for an electron in LAr. The values assume a 100 eV event, but proportions are independent of event energy.}
    \label{fig:Sankey}
\end{figure}

Charged particle events also deposit some thermal energy via sub-excitation electrons; when either the incident particle or secondary electrons are less energetic than it would take to put an electron into the first excitation level, they instead deposit their energy thermally \cite{Chepel_LNG_2013, PLATZMAN_1961}. The proportion of total deposited energy $E_0$ that goes into ionization, excitation, and immediate thermalization is given by \cite{PLATZMAN_1961} as:
\begin{equation}\label{platzman}
    E_0=N_iE_i+N_{ex}E_{ex} +N_i\epsilon ,
\end{equation}
where $\epsilon$ is the average kinetic energy of sub-excitation electrons, $N_i$ and $N_{ex}$ are the relative frequencies of ionization and excitation, and $E_{ex}$ and $E_i$ are the average energies spent to ionize or excite argon atoms. Using the values outlined in Table \ref{tab:energy_summary}, we find that for an incident electron, 26\% of the incident particle energy is deposited thermally, 11\% goes into excitation, and 63\% goes into ionization. With argon scintillation producing 9.8 eV photons, two thirds of the energy that goes into photon-producing states (48\% of total energy) is released as light while the remaining third (26\% of total energy) is released thermally during the various vibrational stages of the de-excitation process. Figure \ref{fig:Sankey} illustrates the full breakdown of energy deposition mechanisms.

\begin{table}[]
    \centering
    \setlength{\tabcolsep}{10pt}
    \caption{Summary of deposition energy contributions to excitation, ionization, and immediate thermalization in LAr.}
    \begin{tabular}{ c | c | c }
        Parameter (Unit) & Value & Reference(s) \\
        \hline
         $E_i$ (eV)   & 15.4 & \cite{DOKE_fanofactors_1976} \\
         $E_{ex}$ (eV)   & 12.7 & \cite{DOKE_fanofactors_1976} \\
         $\epsilon$ (eV)   & 6.3 - 7.7 & \cite{Chepel_LNG_2013} \\
         $N_{ex}/N_i$   & 0.21 & \cite{Segreto_liquidscintillation_2020, DOKE_fanofactors_1976, Chepel_LNG_2013, Kubota_liquidrecomb_1978} \\
         $\tau_f$ (ns)   & 6 $\pm$ 2 & \cite{Hitachi_timedependance_1983} \\
         $\tau_s$ (ns)  & 1590 $\pm$ 100 & \cite{Hitachi_timedependance_1983}
    \end{tabular}
    \label{tab:energy_summary}
\end{table}

\section{\label{sec:Nucleation Model}Bubble Nucleation and the Seitz Model}

The heat spike description of bubble formation (Seitz model) is commonly used as a model of the bubble nucleation process in bubble chamber experiments \cite{Seitz_heatspike_1958, Denzel_MD_2018}. The model outlines the process by which a ``protobubble" forms when sufficient energy is deposited within a local region of a superheated fluid. To form a stable bubble, the total energy deposited must be greater than the heat of vaporization of the liquid plus the surface energy associated with the liquid-gas interface in the bubble. This energy threshold, $Q_{seitz}$, is given by the relation \cite{PICO_bubble}
\begin{equation}
\begin{split}
    Q_{Seitz} \approx 4\pi r_c^2 \left(\gamma - T\frac{\partial \gamma}{\partial T}\right) + \frac{4\pi}{3}r_c^3\rho_b(h_b-h_l)\\ - \frac{4\pi}{3}r_c^3(P_b-P_l)  ,
\end{split}
\end{equation}
where $\gamma$ is the surface tension of the fluid, $\rho_b$ is the density of the bubble, $h_b$ and $h_l$ are the specific enthalpies of the gaseous and liquid states, $P_b$ and $P_l$ are the pressure inside the bubble and of the liquid, and $r_c$ is the critical radius within which energy must be deposited, and is given by the following relationship:
\begin{equation}
    R_c=\frac{2\gamma}{P_b-P_l}.
\end{equation}


Under this model, bubbles are formed when sufficient energy is deposited within a cylindrical track of radius $R_c$ and length $\Lambda R_c$, with $\Lambda=2$ in the original work by Seitz \cite{Denzel_MD_2018}. The model assumes that all thermal energy is deposited immediately into the fluid. As a result, the impacts of time-delayed energy deposition due to scintillation processes (which this work aims to investigate) are not well described.

\section{\label{sec:Simulation Setup}Simulation Setup}

The simulation was run using the HOOMD-blue molecular dynamics package \cite{HOOMDBLUE}. A force-shifted and truncated Lennard-Jones pair potential was applied between the simulated argon atoms as

\begin{equation}
V_\mathrm{LJFS}(r) =
\begin{cases}
V_\mathrm{LJ}(r) - V_\mathrm{LJ}(r_\mathrm{c}), & r \leq r_\mathrm{c}, \\
0, & r > r_\mathrm{c}.
\end{cases}
\end{equation}

\noindent where $r$ is internuclear separation, $r_c$ is a default cutoff radius of 2.5$\sigma$, and $V_\mathrm{LJ}$ is the standard Lennard-Jones potential 
\begin{equation}
\label{eq:LJ_eqn}
    V_{LJ}=4\epsilon\left[\left(\frac{\sigma}{r}\right)^{12}-\left(\frac{\sigma}{r}\right)^6\right].
\end{equation}
Under this potential, forces smoothly approach zero as $r$ approaches $r_c$ \cite{forceshiftedLJ}. The length, energy, and mass scales of the simulation were $\sigma=0.34$ nm, $\epsilon=120K \cdot k_B=10.34$ meV/K, and $m=39.95$ amu. These parameters lead to a characteristic time scale $\tau=\sqrt{\sigma^2m/\epsilon}=$2.152 ps. To verify these parameters in our simulation, it was run at various temperatures in an attempt to induce nucleation with large heat spikes. Below the boiling point of argon ($\sim$87 K), it was not possible to induce boiling with a heat spike, with nucleation behaviour manifesting as soon as the boiling point was crossed. 

\subsection{Preparing the Heat Spike}
Before simulating the particle event, several stages are required to create a superheated state. These steps have been taken from \cite{Denzel_MD_2018, Diemand_homogNuc_2014} and are discussed in greater detail there; here we provide a brief summary of the stages preceding the modified heat spike.

First, simulated particles are initialized in a cubic lattice at a liquid-like density of 0.8 atoms/$\sigma^3$ and ``melted" by assigning random velocities according to the Boltzmann curve for the initial stable liquid state at 85 K. This melted state is allowed to equilibrate over 20,000 steps $dt$, where $dt=0.0025\tau$, and then superheated by raising the temperature past the boiling point of argon over another 20,000$dt$. This superheated state is allowed to stabilize for 20,000$dt$, and then run for 100,000$dt$ under constant-volume, constant-energy (NVE) integration. With the exception of this ``long run" stage, all simulation steps are run at constant temperature and pressure (NPT) or constant pressure and enthalpy (NPH) during the heat spike. Constant volume integration was used for all stages in \cite{Denzel_MD_2018}, which used approximately 14 million particles. However, using NVE for smaller runs of 1 million particles results in sharp  spikes in pressure, disrupting stable bubble formation, due to the low compressibility of liquids. NPT and NPH instead allow the simulation box volume to be rescaled to maintain a constant pressure.

\subsection{Modified Heat Spike and Scintillation}



To model a particle energy deposition in the LAr,  a modified version of the Seitz heat spike model was used. Kinetic energy was imparted to particles within the critical cylinder by increasing their velocities in random directions. The energy deposition of the initial event was spread over $N_\mathrm{spike}=100$ particles. This kept individual speeds down and maintained the stability of the simulation while still modeling the energy transfer from a small number of primary recoils into many secondary recoils, ionizations, and excitations. After the spike, the bubble was allowed to evolve under an NPH ensemble for 300,000$dt$.

The energy transfer steps associated with  scintillation (described in Section \ref{sec:Scintillation in LAr}) were also incorporated into the simulation. The distribution of the primary energy between ionization, excitation, and immediate thermalization processes was taken to be like that for an electron recoil, as described in Equation (\ref{platzman}). Since LAr is transparent to scintillation photons and the time scale for optical photon emission is long compared to the bubble nucleation time, we consider the energy ultimately emitted as scintillation photons to be lost to the nucleation process. To incorporate the energy released thermally as argon dimers form and separate (the delayed thermal energy from Figure \ref{fig:Sankey}), excited state particles were implemented. These excited state particles had the same Lennard-Jones parameters as ground-state argon. A custom updater was then used to probabilistically de-excite the excited states with a specified half-life, changing them back to ground-state and imparting the delayed thermal energy. This results in the initial heat spike being followed by a slower release that follows an exponential time profile as individual particles de-excite.

We abstracted the three-stage relaxation process in ionization events (see Figure \ref{fig:SAr_potentialdiagram}) as a single process with the summed energy release of the three constituent stages and a half-life of 120 ps (55.76$\tau$). As the dimerization (first and third) stages are 1-2 orders of magnitude faster than electron capture (second), this approximation does not significantly change the timing of energy deposition.

In addition to abstracting the de-excitation process as a single step, the simulation assumes that all scintillation results from ionization. In reality, approximately 10\% of the delayed thermal energy comes from direct excitations, which do not need to recapture an electron and as such relax in times on the order of 10 ps, rather than the ionization events which relax with a half-life of 120 ps. Improving this generalization (which impacts less than 2.5\% of total event energy) would require a more in-depth treatment of the kinetics of the de-excitation process.

\subsection{Analysis Methods}
\begin{figure*}
\centering
\begin{subfigure}{0.2\textwidth}
  \includegraphics[width=1.21\linewidth]{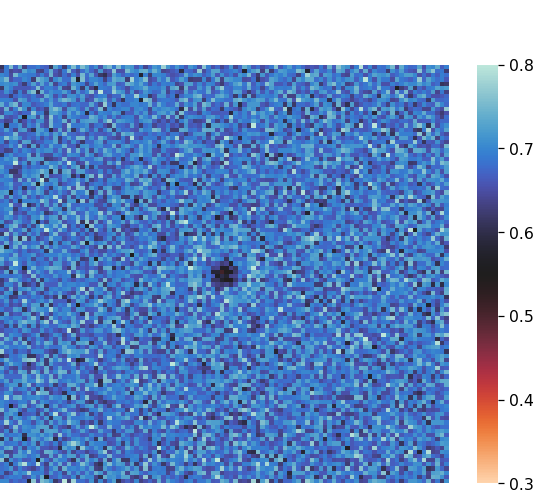}
  \centering
\end{subfigure}%
\begin{subfigure}{0.2\textwidth}
  \includegraphics[width=1.21\linewidth]{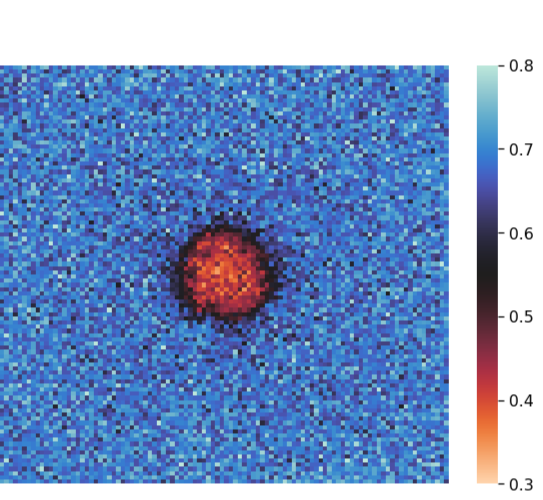}
  \centering
\end{subfigure}%
\begin{subfigure}{0.2\textwidth}
  \includegraphics[width=1.21\linewidth]{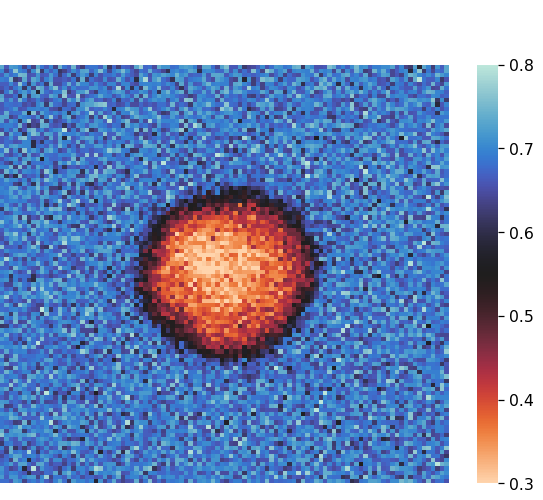}
  \centering
\end{subfigure}%
\begin{subfigure}{0.2\textwidth}
  \includegraphics[width=1.21\linewidth]{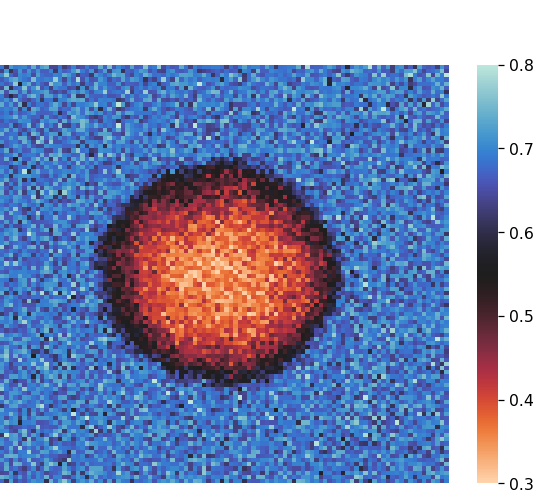}
  \centering
\end{subfigure}%
\\
\begin{subfigure}{0.2\textwidth}
  \includegraphics[width=1.1\linewidth]{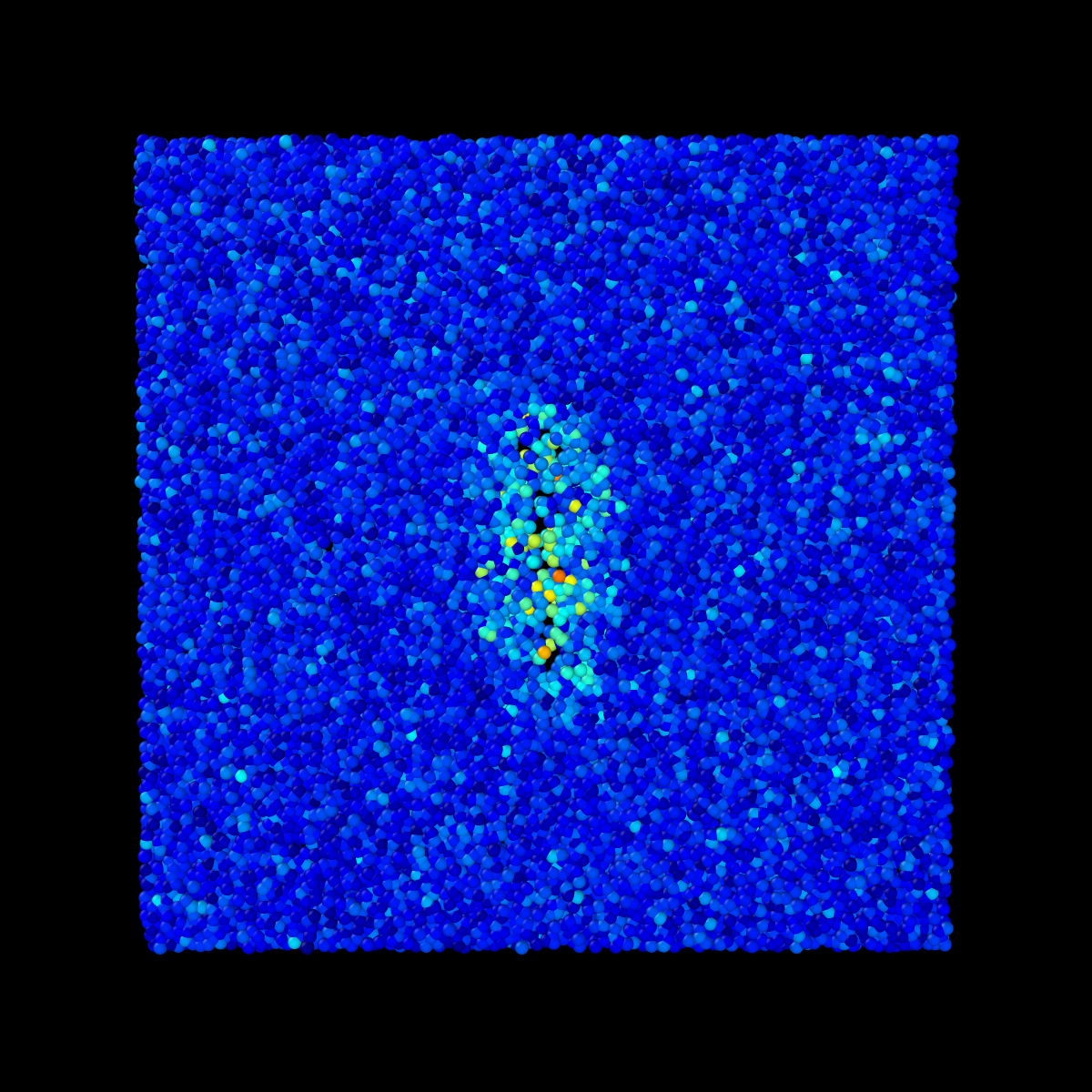}
  \centering
  \caption{t = 250$dt$ (1.35 ps)}
\end{subfigure}%
\begin{subfigure}{0.2\textwidth}
  \includegraphics[width=1.1\linewidth]{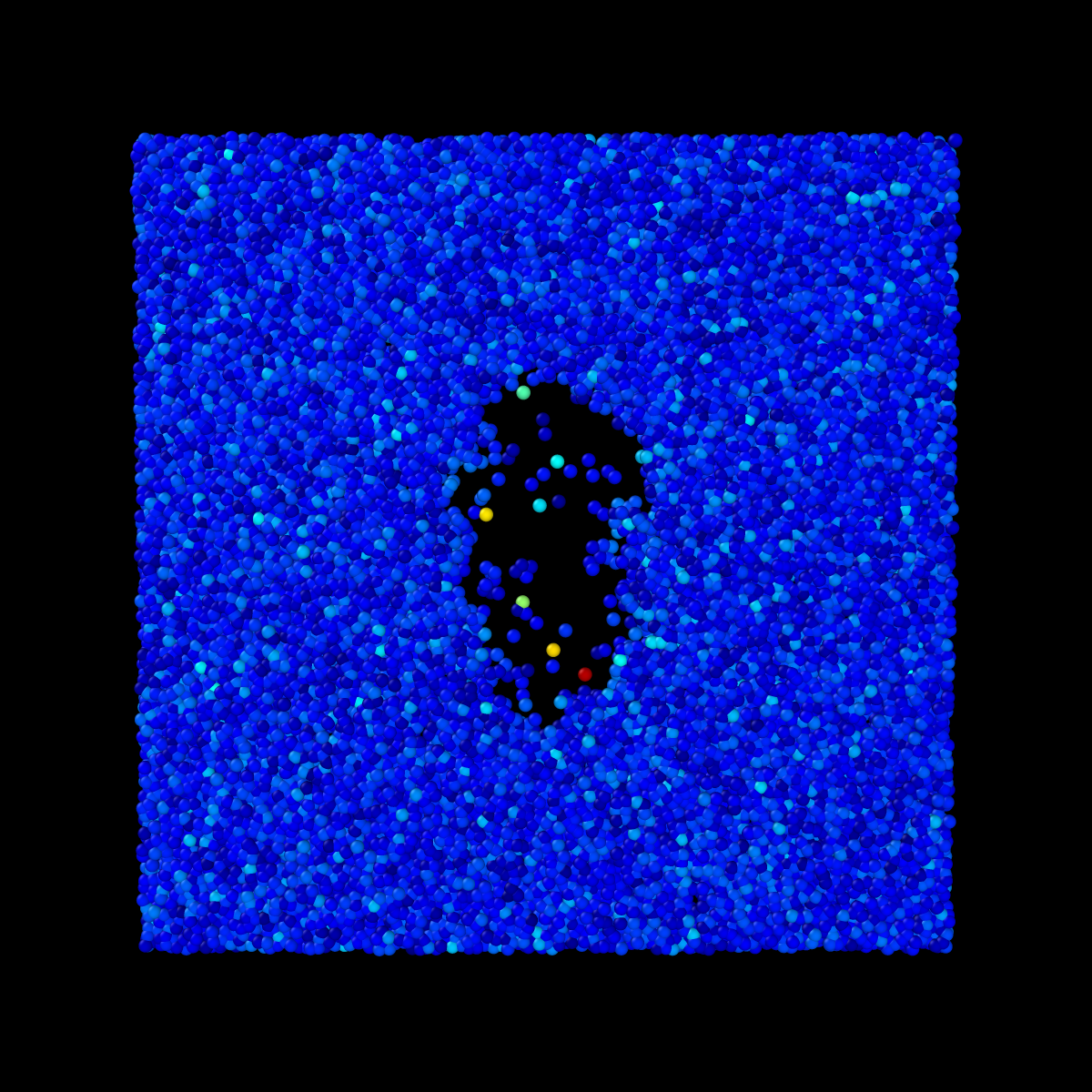}
  \centering
  \caption{t = 34$dt$ (45.73 ps)}
\end{subfigure}%
\begin{subfigure}{0.2\textwidth}
  \includegraphics[width=1.1\linewidth]{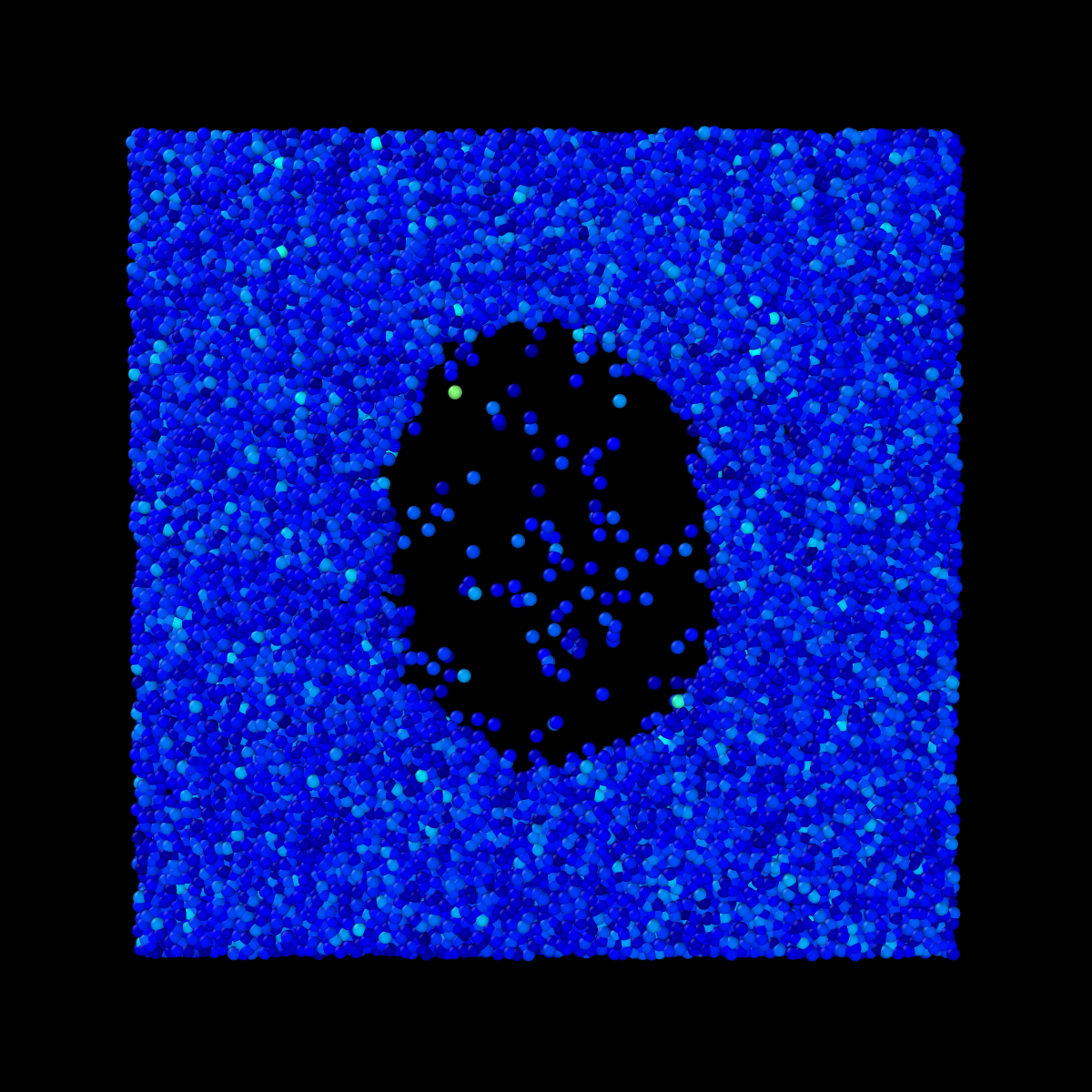}
  \centering
  \caption{t = 100$dt$ (134.5 ps)}
\end{subfigure}%
\begin{subfigure}{0.2\textwidth}
  \includegraphics[width=1.1\linewidth]{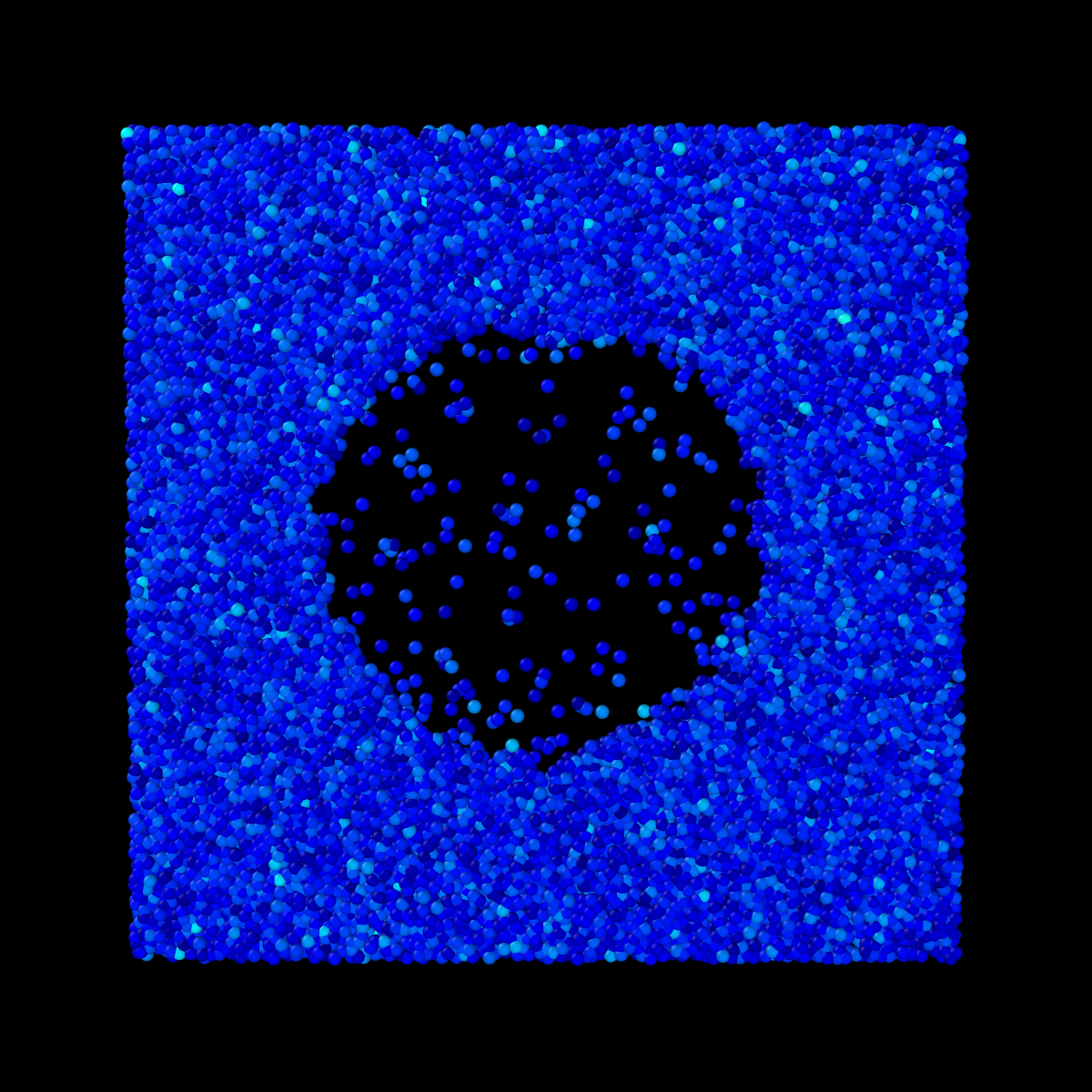}
  \centering
  \caption{t = 1000$dt$ (1345 ps)}
\end{subfigure}%
\caption{\label{fig:nucleation}Evolution of a nucleation event (90K, 920 eV) in multiple views.\vspace{0.25em}
\parbox{\textwidth}{%
\textbf{Top:} Density (atoms$/\sigma^3$) of vertical slices in the x-y plane. \\ \vspace{0.25em}
\textbf{Bottom:} Cross-sectional render in the x-z plane using OVITO \cite{ovito}. Colouring by particle velocity highlights the initial heat spike in the first frame (t = 1.35 ps), and several de-excitations occurring in the second frame (t = 45.73 ps).}
}%
\end{figure*}

Similar to previous work, the simulation box was divided into 100$^3$ cubic voxels and a recursive algorithm was used to identify regions with a density less than 0.2 atoms/$\sigma^3$ \cite{Diemand_homogNuc_2014, Watanabe_cavitation_2010}. This metric was used to track the volume of bubbles over time, while the existence of bubbles anywhere in the simulation was detected by comparing the volume of the entire cell before the heat spike and after letting it evolve for 300,000$dt$. With $N=1,000,000$, simulation cells which grew by more than 10\% tended to contain stably growing bubbles. Those which failed to produce a stable bubble had volume differences of only a few percent, or none at all, depending on how quickly they collapsed back to the liquid state. Figure \ref{fig:nucleation} shows four snapshots of a nucleating bubble, visualized both as density slices and using the OVITO rendering software \cite{ovito}. Vertical slicing illustrates the sphericity of the bubble (as thicker gaseous regions appear brighter), while the colored cross-section highlights energy deposition over time.

While the real SBC detectors are expected to operate at temperatures up to 130 K, the spinodal point for argon as a Lennard-Jones fluid is lower; at a density of 1.1 g/cm$^3$, the spinodal point is 105 K \cite{MALOMUZH_spinodal}. Above this point, the liquid state becomes perturbatively unstable, with small fluctuations causing phase change without a single nucleation point. This behaviour was observed in the simulation, with the entire volume spontaneously boiling above 104 K at 20 psia with an average density of 1.06 g/cm$^3$. As such, this work only examines behaviour in the metastable superheated liquid regime up to 100 K. We note that the discrepancy between the spinodal point of Lennard-Jones argon and the expected run temperatures of the SBC detectors suggests that the use of the Lennard-Jones potential may result in an offset in the bubble nucleation point. However, the relative changes in nucleation point due to scintillation processes, discussed below, should still be valid. 

\raggedbottom

\section{\label{sec:Timescales}Results: Impact of Scintillation on Thresholds}
\begin{figure}
    \centering
    \captionsetup{justification=raggedright, singlelinecheck=false}
    \includegraphics[width=1.1\linewidth]{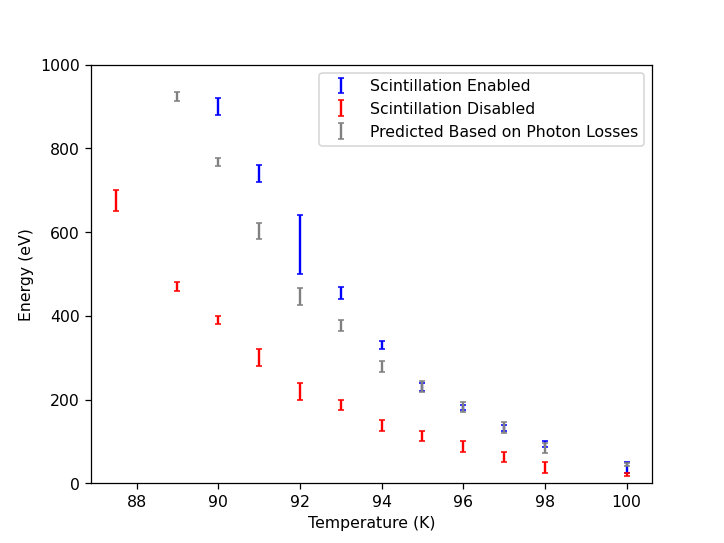}
    \caption{Nucleation energy threshold ranges as a function of temperature at 20 psia. Predicted thresholds for the scintillation points, determined by scaling the ``scintillation disabled" points by the  fraction of energy lost to photon emission, are marked in gray.}
    \label{fig:moneyplot}
\end{figure}

219 simulations (114 with scintillation and 105 without) were run in total, incrementing deposition energy until a bubble nucleated. Temperatures ranged from 85K to 100K, and deposited energies from 10 eV to 1000 eV. In comparing the differences between the simulations with and without scintillation, we consider the fractional difference between the observed thresholds ($Q_{scint}/Q_{non-scint}$). The mean fractional difference over 10 temperature points was 2.16, equivalent to a 53.7\% decrease in the sensitivity to beta events. Figure \ref{fig:moneyplot} illustrates the nucleation thresholds with and without scintillation; note that at 87.5 K and 89 K, even the maximum deposition of 1000 eV was insufficient to cause nucleation with scintillation enabled. 

\subsection{Nucleation Process and Timescale}

The formation of bubbles occurs in two distinct stages in the simulation. In the first stage, which took approximately 250 ps on average, there was a rapid spike in volume as the initial protobubble formed along the deposition track. This occurred regardless of whether or not sufficient energy had been deposited to form a stable bubble.

In the second phase, if enough energy was deposited to form a stable bubble, the growth leveled off to a slower, linear rate. This growth continued indefinitely or until the bubble diameter reached the width of simulation cell, at which point the bubble expanded rapidly due to the disappearance of surface tension. However, this artifact occurred significantly later than the relevant timescales when using more than 100,000 particles. If insufficient energy was deposited and a stable bubble failed to form, the protobubble instead collapsed back into a liquid state.

\begin{figure}
    \centering
    \includegraphics[width=1\linewidth]{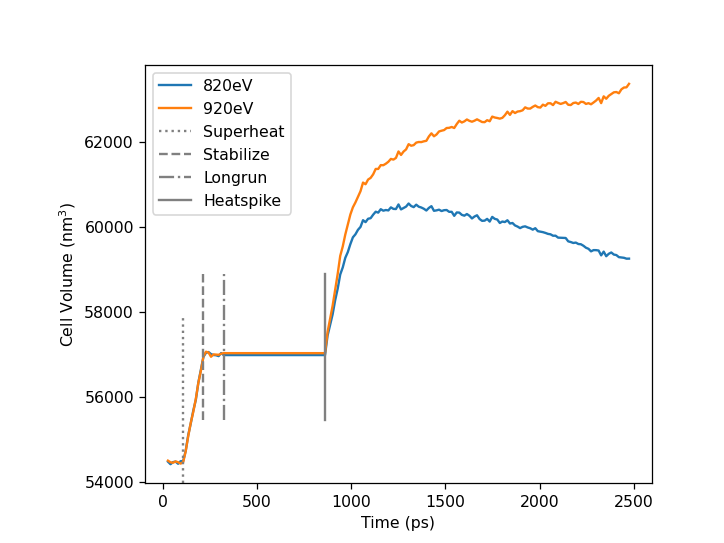}
    \captionsetup{justification=raggedright, singlelinecheck=false}
    \caption{Simulation cell volume over time at 90 K and 20 psia for successful (920 eV) and failed (820 eV) bubble formation with scintillation effects enabled. Beginning of different simulation phases are marked with dashed lines.}
    \label{fig:all_phases}
\end{figure}

This is similar to what was observed in \cite{Denzel_MD_2018}, with a few key differences. They observed an additional stage wherein the bubble grew linearly but was perturbed by volume oscillations indicating a pressure wave propagating through the fluid. They also observed that the initial protobubble would ``overshoot" and then slightly decrease in volume before entering the stable growth regime. We did not observe these behaviors, likely because the presence of the barostat in NPH integration prevented the propagation of a pressure wave through the fluid. Figure \ref{fig:all_phases} shows the simulation cell volume over the entire simulation run, including the preparatory stages.

\subsection{Mechanisms Affecting Nucleation Thresholds}

\begin{figure}
    \centering
    \includegraphics[width=1\linewidth]{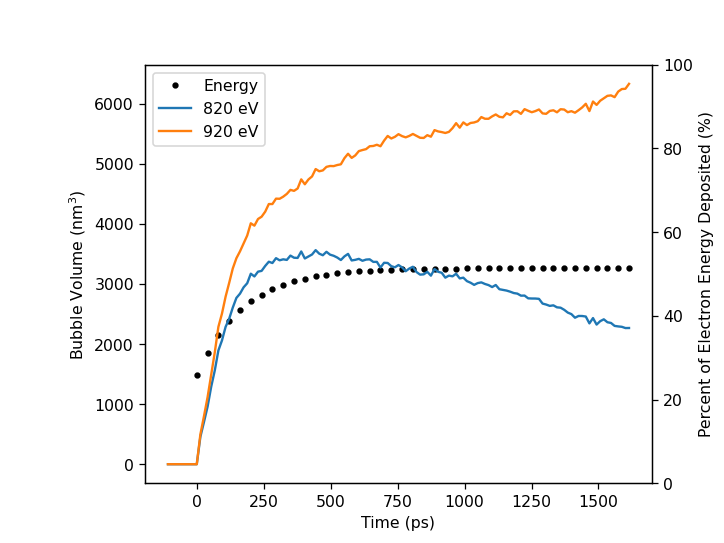}
    \caption{Bubble volume over time at 90 K and 20 psia for stable (920 eV) and collapsing (820 eV) bubble. Proportion of total event energy deposited plotted on same time axis for reference.}
    \label{fig:bubbleWithEnergy}
\end{figure}

There are two mechanisms through which the simulated scintillation effects reduced the sensitivity to electrons. The first is photon losses; the $\sim$50\% of imparted energy that gets converted to photons was simply omitted in the simulation, as xenon-doped LAr is transparent to the scintillation light it produces. The second mechanism, responsible for the rest of the difference in thresholds, is the time-delayed release of ``slow thermal" energy stemming from the formation and separation of dimer states during the relaxation process. As discussed, this release was modeled by the probabilistic de-excitation of 100 initially excited particles with a half-life of 120 ps. Figure \ref{fig:bubbleWithEnergy} shows the evolution of a successful and failed nucleation event, with the proportion of total energy deposited overlaid on the same time axis for comparison.

The expected increase in threshold for events in a scintillating medium relative to events in a non-scintillating medium, can be simply estimated using Equation \ref{eq:estimate_threshold}. 

\begin{equation}\label{eq:estimate_threshold}
    \Delta Q_\mathrm{frac} = \frac{1}{1-E_\mathrm{lost}}.
\end{equation}
As can be seen in Figure \ref{fig:moneyplot}, the observed increase in nucleation threshold for events in the scintillating medium is larger than the expected increase assuming only energy loss to photons. The difference is due to the energy trapped in excited molecular states that is released on time scales longer than those relevant for bubble nucleation. 

With an average time to stable nucleation of approximately 250 ps (116.17$\tau$), $\sim$2.1 excited state half-lives elapsed before bubbles reached stable growth or collapse. As such, 77\% of the delayed thermal energy was deposited in the fluid before bubble growth had stabilized, in addition to the immediate thermal deposition from the heat spike. Equation \ref{eq:estimate_threshold} was used to estimate the fractional threshold differences if all, some, or none of the delayed thermal energy was treated as useful work. Table \ref{tab:estimates} summarizes these estimates. When only the energy deposited before the end of the rapid growth phase is considered, we see close agreement between the estimate and the threshold differences observed in simulation.

\begin{figure}
    \centering
    \includegraphics[width=1\linewidth]{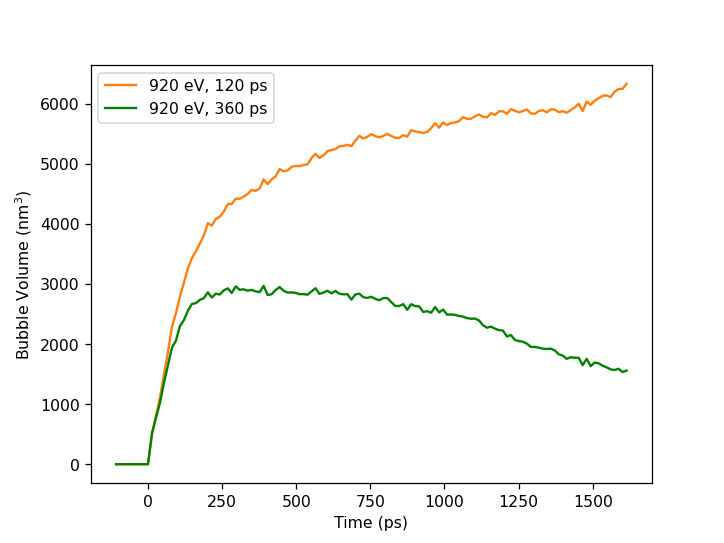}
    \caption{Bubble volumes over time at 90 K and 20 psia for 920 eV event, with exitation half-lives of 120 ps and 360 ps.}
    \label{fig:long_HL}
\end{figure}

An additional simulation with identical parameters to the 920 eV event in Figures \ref{fig:nucleation} and \ref{fig:bubbleWithEnergy} was conducted with the excited state  half-life increased by a factor of 3. Figure \ref{fig:long_HL} shows the bubble growth compared to that of an identical event with the correct excitation half-life of 120 ps. The failure to nucleate, although the total energy and the distribution of that energy was unchanged, further supports the conclusion that energy must be deposited during the initial growth phase to contribute to bubble formation.

\begin{table}[h]
    \centering
    \caption{Estimated threshold differences by amount of delayed thermal energy considered ``lost". Observed threshold increase was by a factor of 2.16.}
    \begin{tabular}{c|c}
        Energy Losses ($E_\mathrm{lost}$) & Predicted Factor Increase\\
        \hline
         All delayed thermal ($E_\mathrm{delayed}$) & 3.86 \\
         $E_\mathrm{delayed}$ after growth phase (23\%) & 2.20 \\
         Only photon losses & 1.94
    \end{tabular}
    \label{tab:estimates}
\end{table}

\vspace{1em}

\section{\label{sec:Conclusions}Conclusions}
Following prior molecular dynamics studies of bubble nucleation in liquid noble gases, we examined the impacts of time-delayed energy release on nucleation thresholds. Qualitatively, the bubble formation process was consistent with observations from other work, with an explosive initial spike in volume followed by linear growth or collapse, depending on whether sufficient energy was deposited. The use of a smaller number of particles under constant-pressure integration likely suppressed the pressure wave from the protobubble, as no volume oscillations were observed in this study. The initial rapid growth phase typically took approximately 250 ps before transitioning into the linear regime.

The minimum nucleation energy for a given temperature was, on average, 2.16 times higher when scintillation was enabled. This is consistent with the increase in threshold that would be expected if the photon energy, as well as any thermal energy released after the rapid growth stage, are considered lost to the local environment. An additional simulation showed that at 90 K, 920 eV was sufficient to nucleate a bubble in LAr with a realistic relaxation half-life of 120 ps, but was insufficient when the half-life was changed to an arbitrary length of 360 ps. This further suggests that energy deposited after the initial growth phase does not affect nucleation (i.e. cannot change a collapsing bubble into a stably growing one) so long as the later events are each small relative to the initial spike. Energy lost as photons appears to be the primary contributor to threshold increases in scintillating liquid-noble bubble chambers relative to non-scintillating equivalents. However, energy lost to time-delayed scintillation processes also removes a meaningful amount of energy from the nucleation process.\newline

Additional work comparing these results to the observed threshold differences in the detector could lay the groundwork to improve this model. As this study did not model the xenon present in the detector ($\sim$100 ppm), a further-refined molecular dynamics model of the interactions between xenon and argon atoms could illustrate how, if at all, xenon doping affects nucleation thresholds.

\begin{acknowledgments}
This research was supported by the Natural Sciences and Engineering Research Council of Canada and the Arthur B. McDonald Canadian Astroparticle Physics Research Institute. We thank the Digital Research Alliance of Canada and Compute Ontario for the computing resources and technical support provided. We also thank Pheerawich Chitnelawong at Queen's University for his consultation and technical advice throughout the course of the project.
\end{acknowledgments}







\newpage
\nocite{*}

\bibliography{bib_main}

\end{document}